\documentclass[twoside,openright,a4paper]{article}
\usepackage{graphicx}
\usepackage{graphics}
\usepackage{epsfig}
\usepackage{amsmath}
\usepackage{amsfonts}
\usepackage{subfigure}
\usepackage{sectsty}
\sectionfont{\large}
.360pk
\font\rt=cmss9.360pk
\font\sd=cmcsc9.360pk

\begin{document}
\centerline{\Large Three-Dimensional Smoothed Particle Hydrodynamics Simulation}
\centerline{\Large for Liquid Metal Solidification Process}

\medskip

\begin{center}
{\sc Raden Ahnaf Faqih S.$^{a,b}$, Christian Fredy Naa$^b$}

\medskip
$^a$Graduate School of Natural Science and Technology, Kanazawa University, Kakuma, Kanazawa 920-1192 Japan, \\
$^b$Faculty of Mathematics and Natural Sciences, Institut Teknologi Bandung, Jl. Ganesha 10, Bandung 40132 Indonesia,\\  
E-mail: radenfaqih@s.itb.ac.id, chris@cphys.fi.itb.ac.id
\\ 
\end{center}

\bigskip

{\parindent 0pt
\textbf{Abstract.}{\em 
The solidification behavior of liquid metal in a container under rapid cooling process is one of the major concerns to be analyzed. In order to analyze its fundamental behavior, a three-dimensional (3D) fluid dynamics simulation was developed using a particle-based method, known as the smoothed particle hydrodynamics (SPH) method. Governing equations that determine the fluid motion and heat transfer involving phase change process are solved by discretizing their gradient and Laplacian term with the moving particles and calculating the interaction with its neighboring particles. The results demonstrate that the SPH mehod can successfully reproduce the behavior and defect prediction of liquid metal solidification process.
}
}

\section{Introduction}
Understanding the solidification behavior of liquid metal is one of the important things for industrial manufacturing field. Sometimes several defects occur if the metal processing is based on the solidification process such as disperse shrinkage, shrinkage cavity, blind cavity, linear cracks, gas porosity, and so on. Various studies are performed for understanding these behaviors. Nowadays, many researchers try to understand it by using numerical study. They make a mathematical model for the multiphase flow, heat transfer, phase transition, and then demonstrate it through a numerical simulation. A review of mathematical modelling of solidification process is discussed in \cite{Hu}. As to the numerical simulation, it can be divided into two methods, the mesh-based method and particle-based method. Conventional mesh-based methods, in general, encounter difficulties in representing multiphase flow. To overcome this problem, the particle-based method which does not need to generate computational mesh can be used. Several particle-based methods have been developed in recent years. One of the particle-based methods which can be applied to simulate incompressible multiphase flows is smoothed particle hydrodynamics. SPH was originally introduced by Lucy \cite{Lucy} and Gingold \& Monaghan \cite{Gingold} to simulate three-dimensional astrophysics problems. However, this method was eventually developed for simulating fluid dynamics and heat transfer problems. SPH method applied to heat transfer for solidification can be seen in \cite{Huppert}. They performed the solidification process for water. In other case, Cleary et al \cite{Cleary} have simulated liquid metal solidification under high pressure die casting. However, they did not explain in detail about the interaction between liquid and solidified metal.

The objectives of this paper are to understand and analyze the application of SPH method for liquid metal solidification. We also want to analyze the behavior and defect prediction of liquid metal especially for the case of rapid solidification process. We first apply the SPH formulation for the fluid flow by attaching a temperature dependence of viscosity. We then apply the SPH method for heat transfer which is based on the enthalpy formulation. To verify that the heat transfer works well, we test the heat transfer model by using 1D static particle. For the phase transition model, we consider the nonisothermal phase change problem. We also consider the interaction between different phases and examine it for 3D simulation case. 

\section{Numerical model}
\subsection{The SPH method}
In SPH, the fluid is represented by a set of particles that can move freely. Each particle carries some fundamental physical properties, such as mass, position, velocity, density, and any other related properties. The value of any function $f$ at a particle can be approximated by summing over the properties of its neighboring particles. The SPH interpolation of the function $f$ of particle $i$ at position $\textbf{r}_i$ is approximated by:
\begin{equation}
f(\textbf{r}_i)\approx\sum_j \frac{m_j}{\rho_j}f_jW(\textbf{r}_i-\textbf{r}_j,h),
\label{eq:1}
\end{equation}
where index $j$ corresponds to any neighboring particle of particle $i$, $m_j$ and $\rho_j$ are the mass and the density for particle $j$, $f_j$ is the value of $f$ for particle $j$, the function $W$ is an interpolation smoothing kernel, and $h$ is a smoothing length that defines the radius of influence around the particle $i$. 

There are many kinds of smoothing kernel functions in the literature and here we choose cubic spline kernel function as below \cite{Liu}:
\begin{equation}
W_{ij} = W(\textbf{r}_{ij},h) = C
  \begin{cases}
   \frac{2}{3}- \left(\frac{|\textbf{r}_{ij}|}{h}\right)^2 + \frac{1}{2}\left(\frac{|\textbf{r}_{ij}|}{h}\right)^3 & \text{, } 0 \leq \frac{|\textbf{r}_{ij}|}{h} < 1 \\
   \frac{1}{6}\left(2-\left(\frac{|\textbf{r}_{ij}|}{h}\right)\right)^3 & \text{, } 1 \leq \frac{|\textbf{r}_{ij}|}{h} < 2 \\
   0       & \text{, } 2 \geq \frac{|\textbf{r}_{ij}|}{h}
  \end{cases}
\label{eq:2}
\end{equation}
where $\textbf{r}_{ij} = \textbf{r}_i-\textbf{r}_j$ and $C = 3/(2\pi h^2) $ for 3D simulation. The radius of influence of this function is $2h$. It means that the particle $i$ will not be affected by any neighboring particles farther than $2h$. However, the closer distance of the neighboring particles to the particle $i$, the greater will be the influence accepted by the particle $i$.

The derivative of the function $f$ is obtained by differentiating the interpolation Eq.(\ref{eq:1}) which is given by:
\begin{equation}
\nabla f(\textbf{r}_i)\approx\sum_j \frac{m_j}{\rho_j}f_j\nabla_i W_{ij}
\label{eq:3}
\end{equation}
where
\begin{equation}
\nabla_i W_{ij} = \nabla W(\textbf{r}_{ij},h) = \frac{1}{h} \frac{\textbf{r}_{ij}}{|\textbf{r}_{ij}|}F(|\textbf{r}_{ij}|, h) \nonumber
\end{equation}
 and $F$ is the derivative of the kernel function Eq.(\ref{eq:2}).

\subsection{Fluid model}
The governing equations of fluids in SPH method are based on the Navier-Stokes equations in the Lagrangian form. The basic equations are given by \cite{Wendt}:
\begin{eqnarray}
\frac{d\rho}{dt} &=& -\rho\nabla\cdot \textbf{v}\\
\label{eq:4}
\rho\frac{d\textbf{v}}{dt} &=& -\nabla p + \mu\nabla^2\textbf{v}+\textbf{F}
\label{eq:5}
\end{eqnarray}
where $t$ is time, $\textbf{v}$ is velocity, $p$ is pressure, $\mu$ is viscosity, and $\textbf{F}$ is external force. Eq.(4) is known as the continuity equation which describes the evolution of the fluid density over time. Eq.(\ref{eq:5}) is known as the momentum equation which describes the acceleration of the fluid. The first term on the right hand side involving the pressure gradient corresponds to the pressure force and the second term involving viscosities corresponds to the viscous force. The external forces used here are gravitational force and repulsive force from the boundaries.

By employing the SPH interpolation of Eq.(\ref{eq:3}) to the Eq.(4) and Eq.(\ref{eq:5}) then the SPH representation of the continuity and the momentum equation can be written as \cite{Liu}:
\begin{eqnarray}
\frac{d\rho_i}{dt} &=& \sum_j m_j (\textbf{v}_i - \textbf{v}_j)\cdot\nabla W_{ij},\\
\label{eq:6}
\frac{d\textbf{v}_i}{dt} &=& -\sum_j m_j \left[\left(\frac{P_i}{\rho_i^2} + \frac{P_j}{\rho_j^2}\right)-\frac{\xi}{\rho_i \rho_j}\frac{4\mu_i\mu_j}{\left(\mu_i + \mu_j\right)}\frac{\textbf{v}_{ij}\textbf{r}_{ij}}{\textbf{r}_{ij}^2 +\eta^2}\right]\nabla_i W_{ij} + \textbf{F}.
\label{eq:7}
\end{eqnarray} 
The viscous force used here is the laminar viscosity term which was introduced by Morris and Monaghan \cite{Morris} with $\mu_i$ and $\mu_j$ the dynamic viscosity of the fluid corresponding to particles $i$ and $j$. Further, $\xi = 4.963$ is a viscous scaling factor \cite{Cleary1} and $\eta$ is a small parameter to prevent singularity when $\textbf{r}_{ij}$ goes to zero (usually $\eta \sim 0.1h$).

An equation of state is required to calculate the pressure in Eq.(\ref{eq:7}). The equation of state used here is quasi-compressible form which is calculated by using the density information from Eq.(6). The equation of state is given by \cite{Cleary1}:
\begin{equation}
P = \beta \left[\left(\frac{\rho}{\rho_0}\right)^\gamma -1\right] 
\label{eq:EOS}
\end{equation}
where $\beta = c^2\rho_0 / \gamma$ is the magnitude of the pressure, $\rho_0$ is the reference density, $\gamma = 7$ for liquid metal, and $c$ is the speed of sound. The value of speed of sound should be large enough  to ensure that the density fluctuation is less than 1\% or close to the incompressible flow.

\subsection{Heat transfer and phase transition model}
The heat transfer process occurs by cooling down the system through the boundaries. Here, the boundaries are modeled by a set of SPH particles which have a certain temperature and thermal conductivity. The temperature and thermal conductivity of the boundary particles are assumed to be constant during the solidification process. We also assume the boundary as adiabatic boundary where there is no flux to the ambient environment. The heat exchange between the particles (liquid, solid, and boundary) occurs by considering the heat conduction. The model of heat conduction is based on the enthalpy method which is given by:
\begin{equation}
\frac{dH}{dt} = \frac{1}{\rho}\nabla (k\nabla T) \nonumber
\end{equation}
where $H$ is enthalpy, $k$ is thermal conductivity, and $T$ is temperature. 

The SPH formulation of this equation is approximated by using the modified SPH approximation for second derivative which is developed by Cleary and Monaghan \cite{Cleary2}:
\begin{equation}
\frac{dH_i}{dt} = \sum_j \frac{m_j}{\rho_i\rho_j}\frac{4k_i k_j}{(k_i + k_j)}(T_i - T_j)\frac{\textbf{r}_{ij}\cdot\nabla_i W_{ij}}{\textbf{r}_{ij}^2 + \eta^2}.
\label{eq:8}
\end{equation}
This equation ensures that the heat flux is automatically continuous across the different material interfaces, such as between the mold and liquid metal. This also allows multiple phases with different conductivities to be accurately simulated \cite{Cleary2}.

\begin{figure}[h]
\centering
\includegraphics[width=2in]{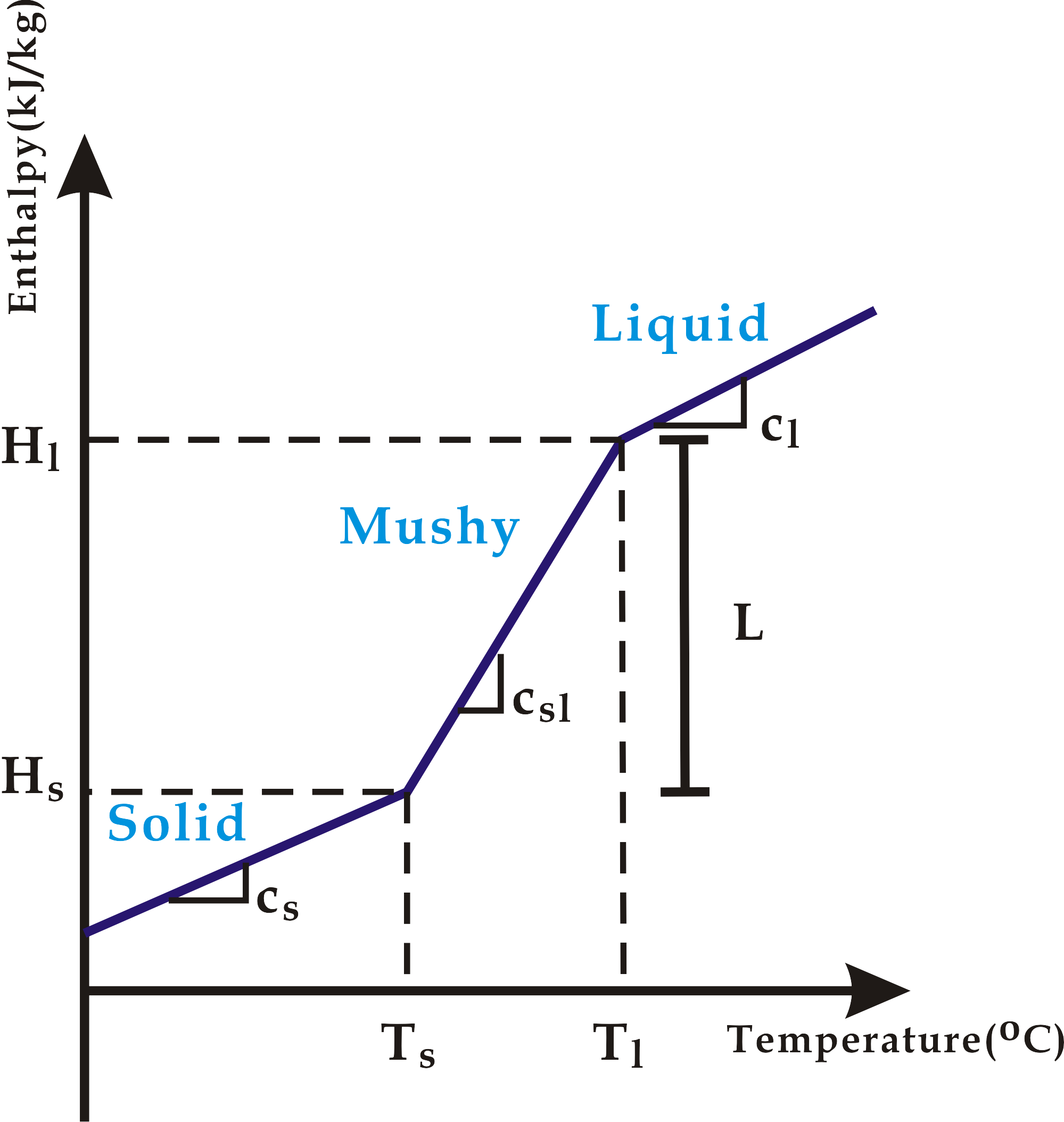}
\caption{Relationship between enthalpy and temperature for nonisothermal phase change}
\label{fig:mushy}
\end{figure}

The temperature in Eq.(\ref{eq:8}) is calculated by considering the relationship between enthalpy and temperature. In the liquid metal solidification case, the phase change occurs over a range of temperature. The phase transition goes through an intermediate phase which is known as mushy phase. So, the nonisothermal phase change is involved to correlate the enthalpy and temperature. The illustration of the nonisothermal phase change is shown in Figure \ref{fig:mushy}. This relation ensures that the temperature is continuous across the different phases. 

The relationship between enthalpy and temperature can be written as \cite{Nedjar}:
\begin{equation}
H = 
  \begin{cases}
   \int_{T_{ref}}^T c_s(T)dT & \text{, } T \leq T_s \text{ $~~~~~~~$ (Solid)}, \\
   \int_{T_{ref}}^{T_s} c_s(T)dT + \int_{T_s}^T \frac{\partial L}{\partial T}dT & \text{, } T_s < T \leq T_l \text{$~~$(Mushy)},\\
   \int_{T_{ref}}^{T_s} c_s(T)dT + L + \int_{T_l}^T c_l(T)dT & \text{, } T > T_l \text{$~~~~~~~~$ (Liquid)},
  \end{cases}
  \label{eq:9}
\end{equation}
where $T_{ref}$ is the reference temperature, $T_l$ and $T_s$ are respectively the solidus and liquidus temperatures, and $c_l$ and $c_s$ are the specific heats in the solid and liquid phases, respectively. If we consider the specific heats to be constant then Eq.(\ref{eq:9}) becomes:
\begin{equation}
H = 
  \begin{cases}
   c_s T & \text{, } T \leq T_s, \\
   c_s T_s + c_{sl}(T-T_s)& \text{, } T_s < T \leq T_l,\\
   c_s T_s + c_{sl}(T_l - T_s) + c_l(T-T_l) & \text{, } T > T_l,
  \end{cases}\nonumber
\end{equation}

The temperature of each particle is then calculated by using this relation in the temperature term:
\begin{equation}
T_i = 
  \begin{cases}
   \frac{H_i}{c_s} & \text{, } H_i \leq H_s = c_s T_s \text{ $~~~~~~~~~~~~~~~~~~~~~~~~$ (Solid)}, \\
   T_s + \frac{H_i - H_s}{c_{sl}} & \text{, } H_s < H_i \leq H_l = H_s + c_{sl} (T_l - T_s) \text{$~$(Mushy)}, \\
   T_s + \frac{H_i - H_l}{c_l} & \text{, } H_i > H_l \text{$~~~~~~~~~~~~~~~~~~~~~~~~~~~~~~~~~~~$ (Liquid)}, 
  \end{cases} \nonumber
\end{equation}
where $c_{sl} = L/(T_l - T_s)$. The latent heat $L$ describes the energy released by a particle to change the phase from liquid to solid.

\subsection{Liquid-solid interaction model}
\begin{figure}[h]
       \centering
       \subfigure[]{
                \centering
                \includegraphics[width=0.8in]{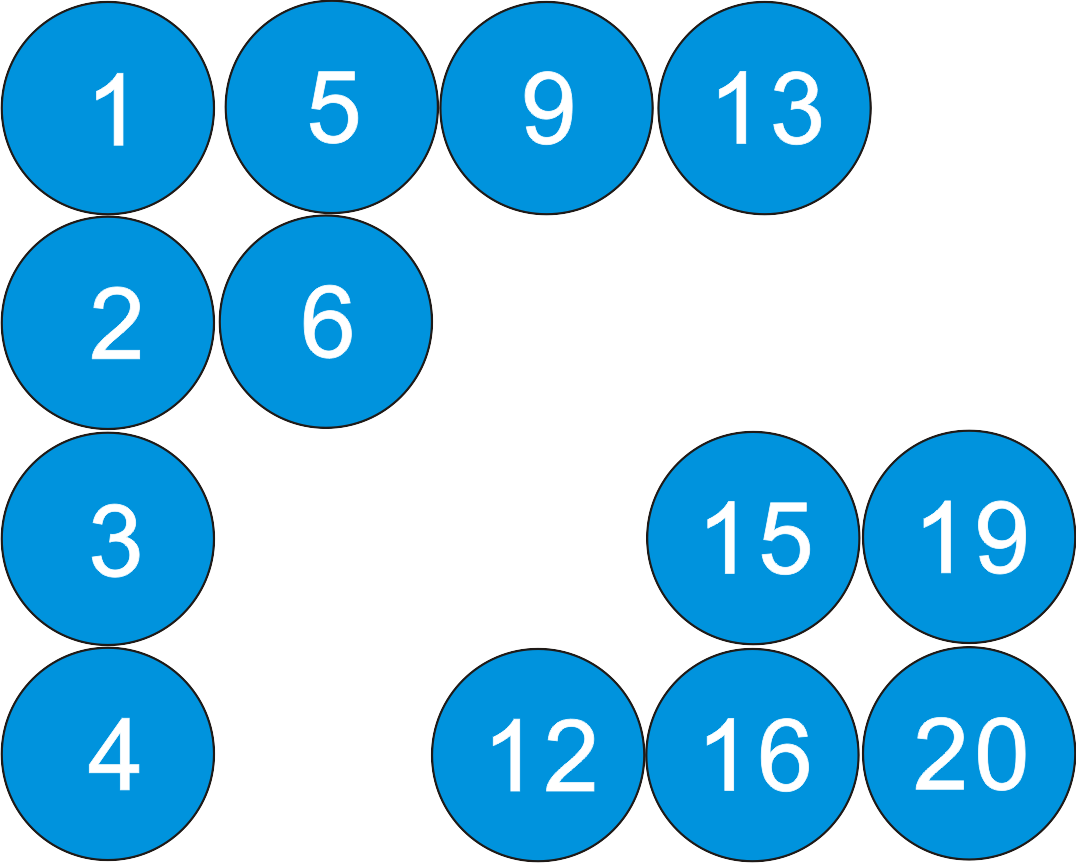}
                \label{fig:solidA}
        }
        \quad
        \subfigure[]{
                \centering
                \includegraphics[width=0.8in]{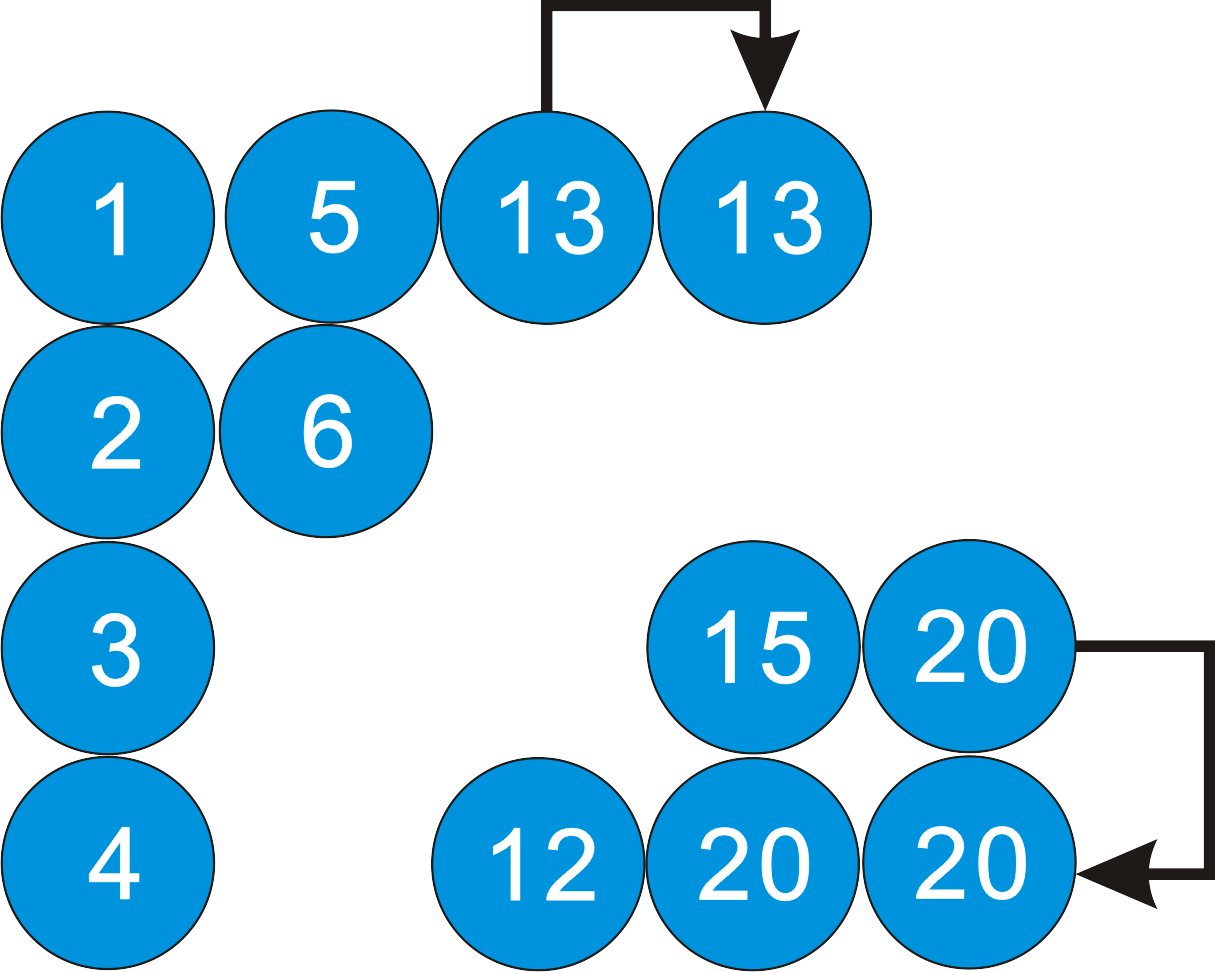}
                \label{fig:solidB}
        }
        \quad        
        \subfigure[]{
                \centering
                \includegraphics[width=0.8in]{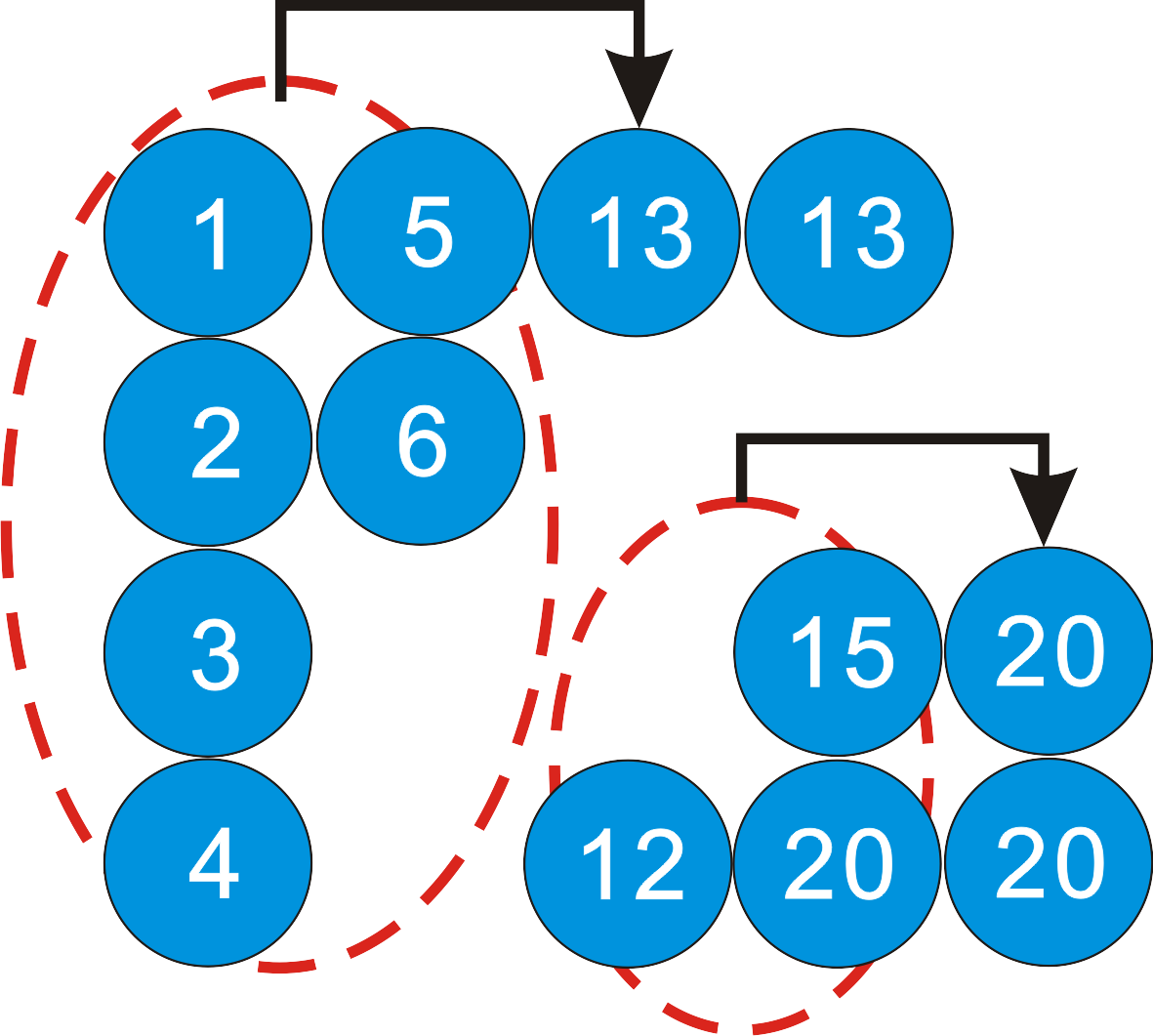}
                \label{fig:solidC}
        }
        \quad       
        \subfigure[]{
                \centering
                \includegraphics[width=0.8in]{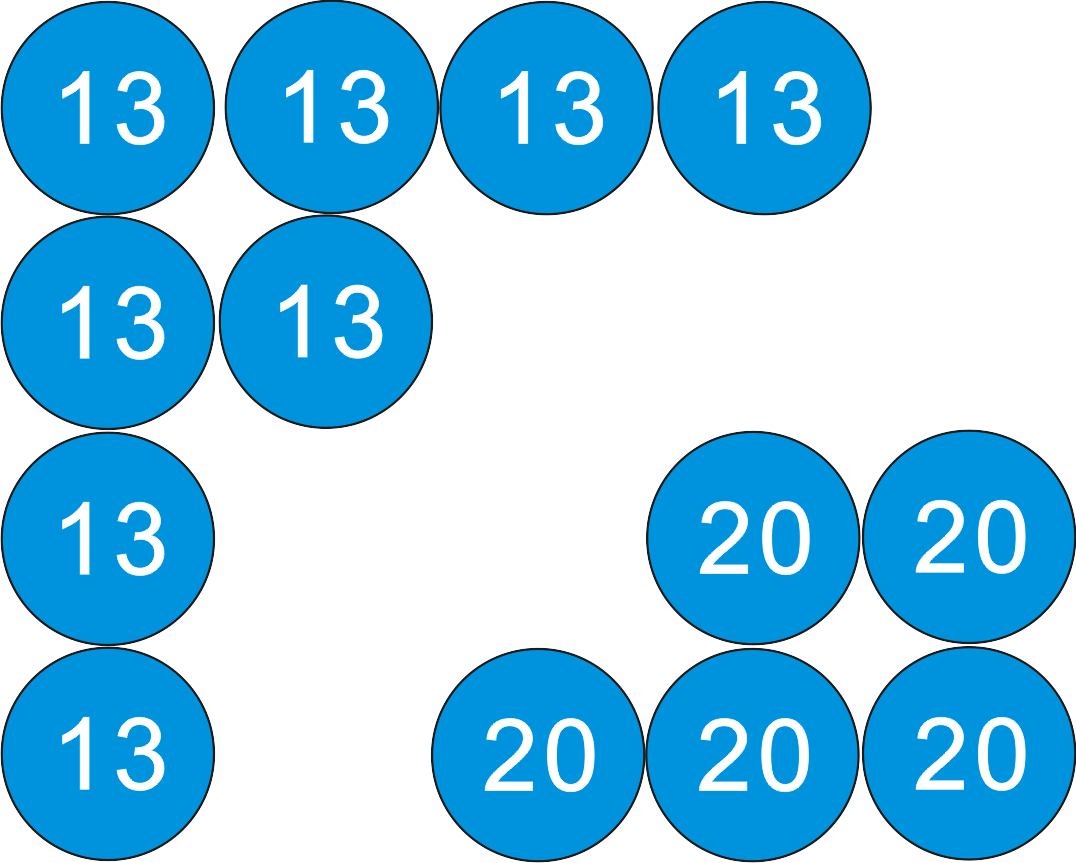}
                \label{fig:solidD}
        }        
        \caption{Illustration of calculating the number of solid groups}\label{fig:solidGroup}
\end{figure}

The phase transition of a particle from liquid to solid occurs when the particle temperature is below the solidus temperature. Here, we modelled the solid particles as a viscous pseudo-fluid. It means that the behavior of solid particles is like a fluid (liquid or mushy) particles but they move under high viscosity. This approach is used to keep the forces acting on the solid particles and fluid particles maintained. If the solid particle has more than two of its neighbors being solid phase then they act and move together as a solid group (rigid body). The illustration how to create the solid group is shown in Figure \ref{fig:solidGroup}. First, each solid particle is assigned an initial index, as shown in Figure \ref{fig:solidA}. After that, each particle index is updated by the maximum index of its neighbor particle (Figure \ref{fig:solidB}). This process is repeated until each solid particle index converges to the maximum index in each group (Figure \ref{fig:solidC}). After some steps, we will only have the maximum index on each group and it means the number of the solid group is determined by how many remaining indices we have. For example in the Figure \ref{fig:solidD}, we have two separate solid groups because finally we have two indices that are 13 and 20.

Here, we consider the solid group as an undeformable object. In order to prevent the deformation, the solid particle's movement should be converted to the translational and rotational motion related to the center of mass of its solid group. The center of mass is calculated by using the following equation:
\begin{equation}
\textbf{r}_c = \frac{1}{N}\sum_j \textbf{r}_j \nonumber
\end{equation}
where $N$ is the number of solid particle in the solid group.

The translational and rotational motion is considered by calculating the moment of inertia and angular velocity of the solid group. The moment of inertia, $I$, and angular velocity, $\omega$, is calculated by the following equations:
\begin{equation}
I = \sum_i |\textbf{q}_i|^2 
\; \text{and }
\omega = \frac{1}{I}\sum_i \textbf{q}_i \times \textbf{v}_i. \nonumber
\end{equation}  
Here, $\textbf{q}_i = \textbf{r}_i - \textbf{r}_c$ is the relative position of particle $i$ to the center of mass. 

During the simulation, the center of mass and 
the rotation matrix are updated by using the following equations:
\begin{equation}
\textbf{r}_c(t + \Delta t) = \textbf{r}_c(t) + \Delta t \textbf{v}_c\left(t+0.5\Delta t\right) 
\; \text{and }
\textbf{R}(t + \Delta t) = \textbf{R}(t) + \Delta t \textbf{Z}\left(t+0.5\Delta t\right) \nonumber
\end{equation}
where $\textbf{v}_p$ is the center of mass velocity and the angular velocity tensor $\textbf{Z}(t)$ in is given by:
\begin{equation}
\textbf{Z}(t) = 
\left[ 
\begin{array}{ccc}
0 & -\omega_z & \omega_y \\
\omega_z & 0 & -\omega_x \\
-\omega_y & \omega_x & 0 \\
\end{array}
\right]\textbf{R}(t). \nonumber
\end{equation}
The rotation matrix in 3D coordinates system is given by:
\small
\begin{equation}
\textbf{R}_x(\theta) = 
\left[ 
\begin{array}{ccc}
1 & 0		 	& 0 			\\
0 & \cos\theta 	& \sin\theta 	\\
0 & -\sin\theta & \cos\theta 	\\
\end{array}
\right], 
\; 
\textbf{R}_y(\theta) = 
\left[ 
\begin{array}{ccc}
\cos\theta 	& 0 		& -\sin\theta 	\\
0 			& 1 		& 0			 	\\
\sin\theta 	& \omega_x 	& \cos\theta 	\\
\end{array}
\right],
\;
\textbf{R}_z(\theta) = 
\left[ 
\begin{array}{ccc}
\cos\theta 	& \sin\theta 	& 0 \\
-\sin\theta & \cos\theta 	& 0 \\
0 			& 0 			& 1 \\
\end{array}
\right]. \nonumber
\end{equation}
\normalsize
After we get the angular velocity and the center of mass, the relative velocity and position of each solid particle to the center of mass can be written as below:
\begin{eqnarray}
\textbf{v}_i(t+\Delta t) &=& \textbf{v}_c(t) + \omega \times \textbf{q}_i(t), \nonumber\\
\textbf{r}_i(t+\Delta t) &=& \textbf{R}(t+\Delta t)\left[\textbf{r}_i(t) - \textbf{r}_c(t)\right] + \textbf{r}_c(t+\Delta t). \nonumber
\end{eqnarray}

\section{Numerical simulations}
\subsection{One-dimensional sinusoidal temperature}
We simulate the cooling down process in 1D case to verify that the SPH method for heat conduction works well. We use 1m block with the initial temperature distribution is $T(x,0)=\sin\pi x$. The temperature on the boundaries are constant. We set the boundary temperature to 0$^o$C. The exact solution of this problem is
\begin{equation}
T(x,t)=e^{\pi^2\alpha t} \sin\pi x \nonumber
\end{equation}
where $\alpha = k/\rho c_p = 0.1$.
\begin{figure}[h]
\centering
\includegraphics[width=2.9in]{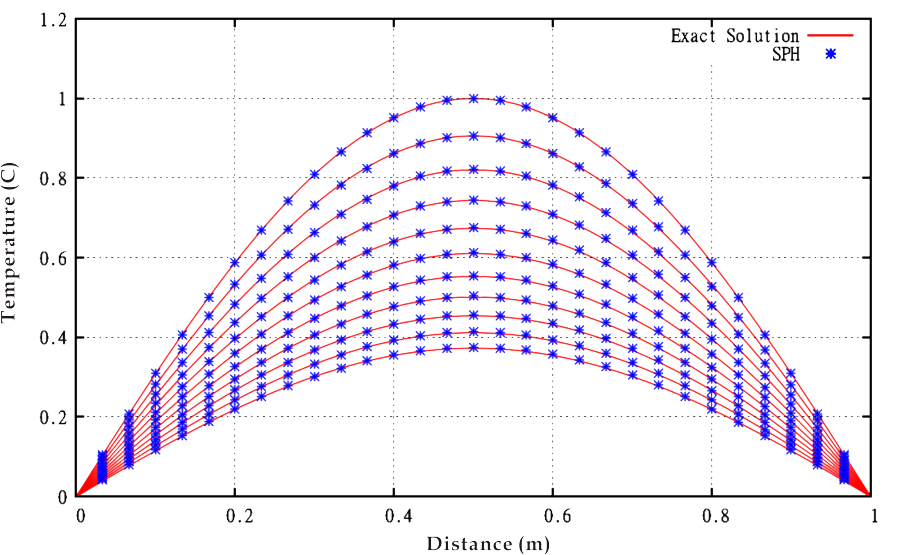}
\caption{Heat conduction for 1D sinusoidal temperature}
\label{fig:sinusoidal}
\end{figure}

Figure \ref{fig:sinusoidal} shows the SPH solution for heat conduction fitted with the exact solution for times $t=0,0.1,0.2,...,1$. The result shows that the constant temperature on the boundaries acts as heat sink of the system.

\subsection{Three-dimensional liquid metal solidification}
\subsubsection{Physical set up and material properties}
The materials used in this simulation are liquid aluminium as the liquid metal and steel as the mold.
The geometry of the mold steel is cube-shaped with the dimension 0.3m$\times$0.3m$\times$0.3m. The temperature of the mold steel is 200$^o$C and the thermal conductivity is $30$ W/mK.

\begin{figure}[h]
       \centering
       \subfigure[]{
               \centering
               \includegraphics[width=2in]{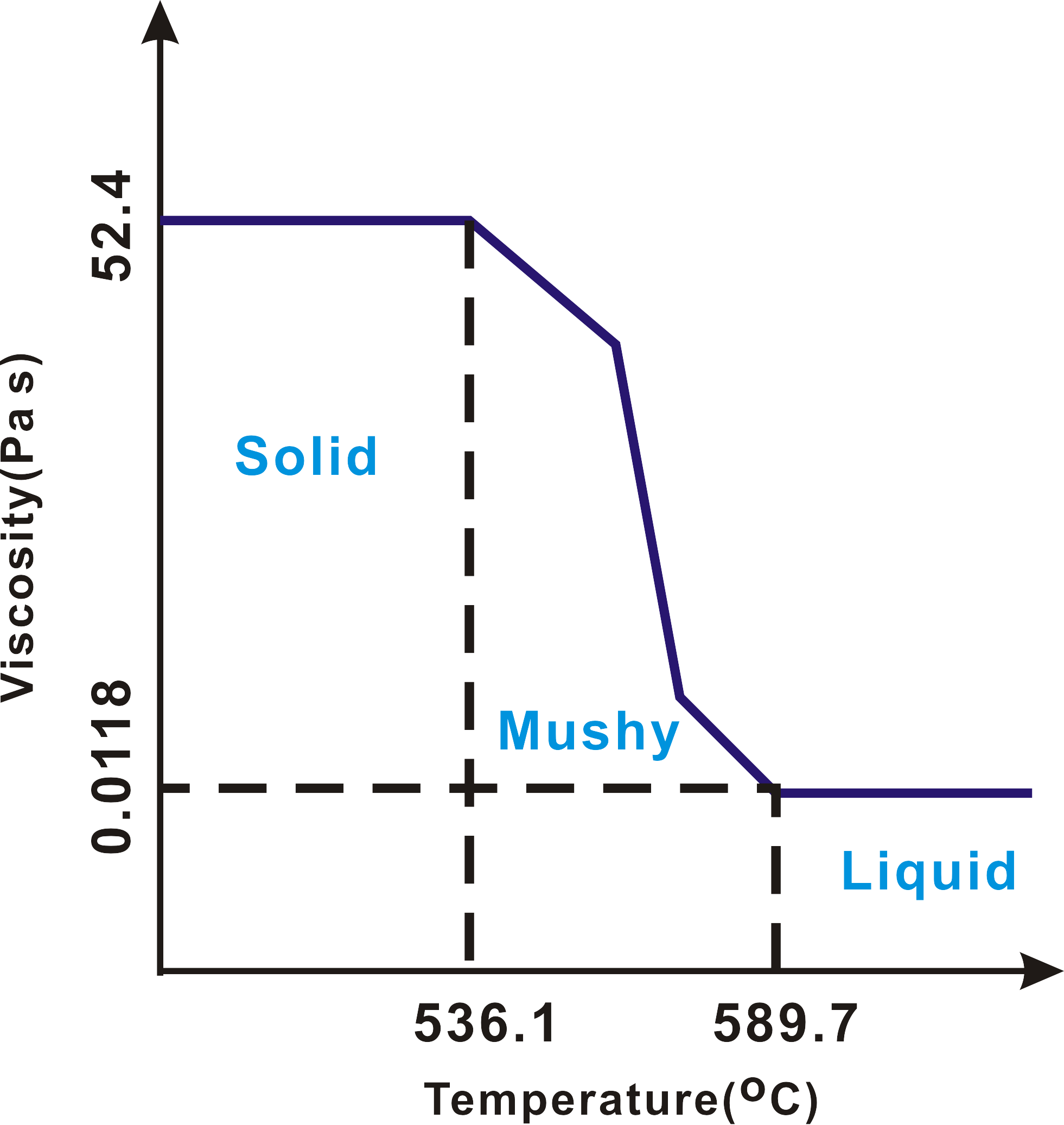}
               \label{fig:viscosityVStemperature}
       }
       \qquad 
       \subfigure[]{
                \centering
                \includegraphics[width=2in]{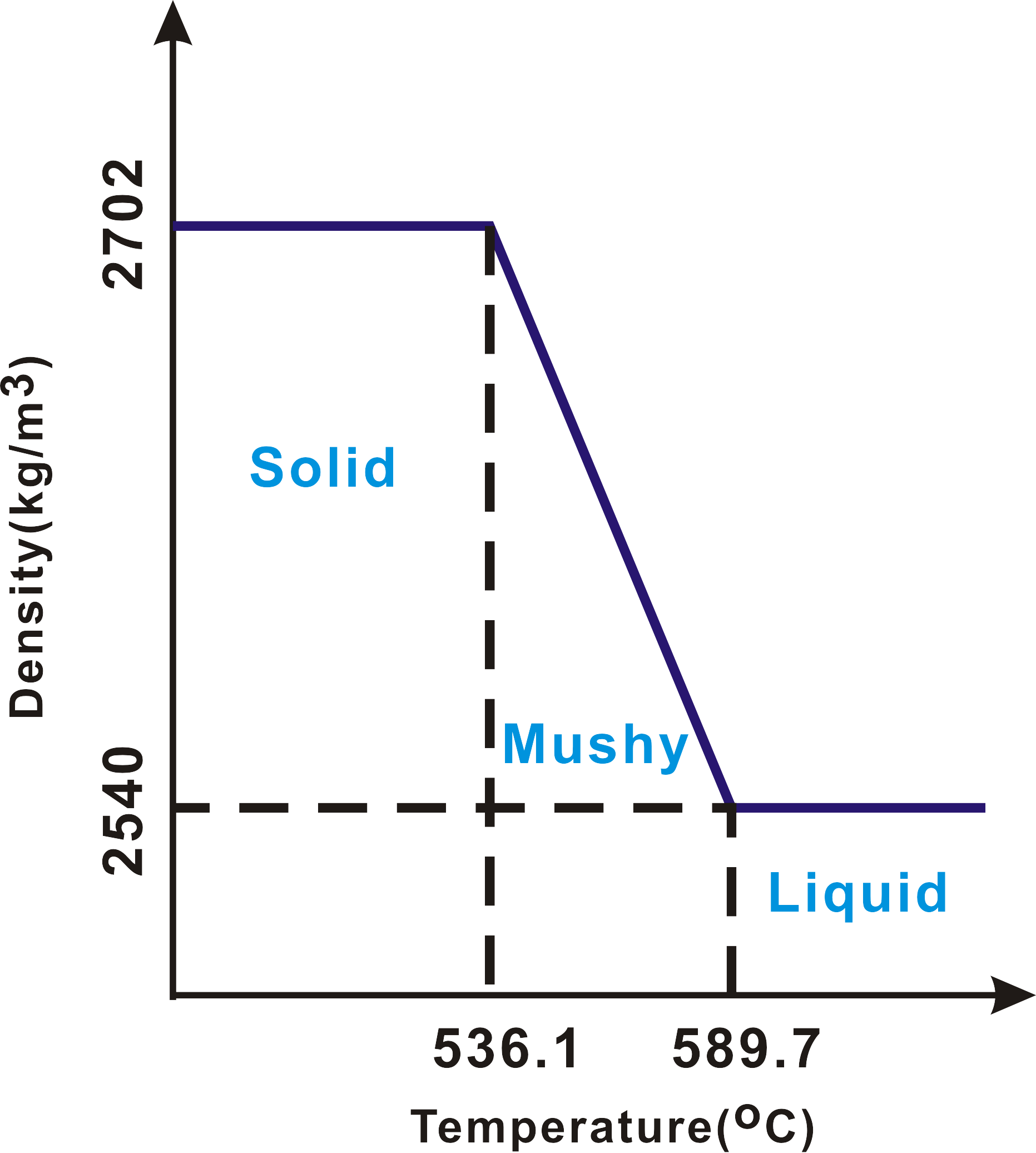}
                \label{fig:densityVStemperature}
        }
        
        \caption{Relation of material properties with temperature: (a) viscosity and (b)  density }\label{fig:materialProperties}
\end{figure}

The mold is fully filled by the liquid aluminium. The liquid aluminium data is taken from \cite{Cleary3}. The solidus and liquidus temperatures are 536.1$^o$C and 589.7$^o$C respectively. The viscosity of the liquid aluminium is temperature dependent. The variation of the viscosity with temperature is shown in the Figure \ref{fig:viscosityVStemperature}. The viscosity above the liquidus temperature is 0.0118 Pa s and the viscosity below the solidus temperature is 52.4 Pa s. The density of the liquid aluminium is also temperature dependent. The liquid metal has a density of 2540 kg/m$^3$ and the solid metal has a density of 2702 kg/m$^3$. In the mushy zone, the variation of density with temperature is close to linear. The variation of the density is shown in the Figure \ref{fig:densityVStemperature}. The other data for the aluminium properties can be seen in Table \ref{table:1}.

\begin{table}[h]
\caption{Thermal physical properties of aluminium}
\centering
\begin{tabular}{l l l}
\hline
Parameter & Value & Unit \\
\hline
Thermal conductivity of liquid & 94.14 & W/mK\\
Thermal conductivity of solid & 237.65 & W/mK\\
Specific heat of liquid & 1.084 & kJ/kg/K\\
Specific heat of solid & 0.963 & kJ/kg/K\\
Initial temperature & 600 & $^o$C\\
Latent heat & 398 & kJ/kg\\
\hline
\label{table:1}
\end{tabular}
\end{table}

\subsubsection{Results and discussions}
\begin{figure}[h]
       \centering
       \subfigure[]{
               \centering
               \includegraphics[width=1.78in]{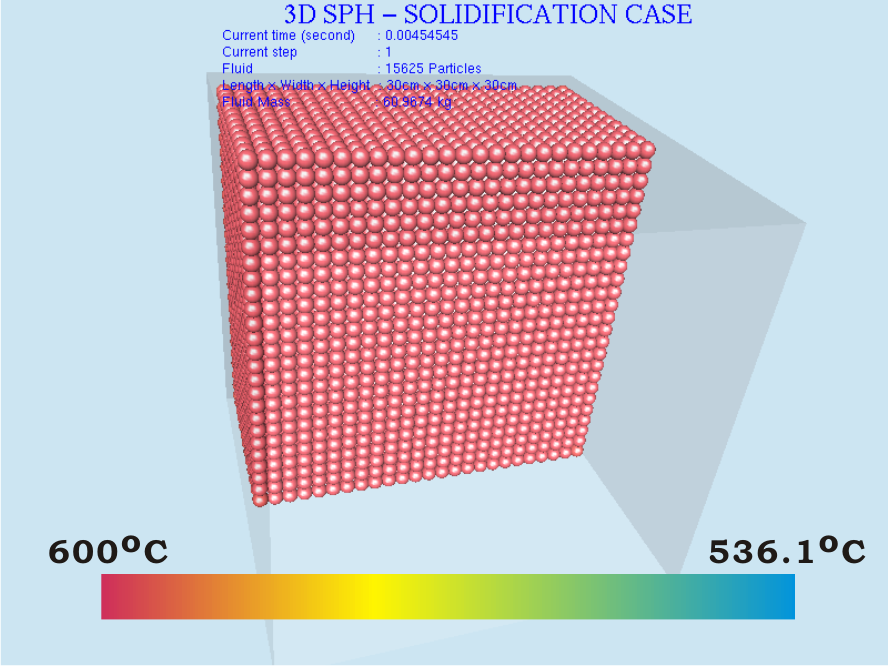}
               \label{fig:Result1}
       }
       \;      
	   \subfigure[]{
               \centering
               \includegraphics[width=1.78in]{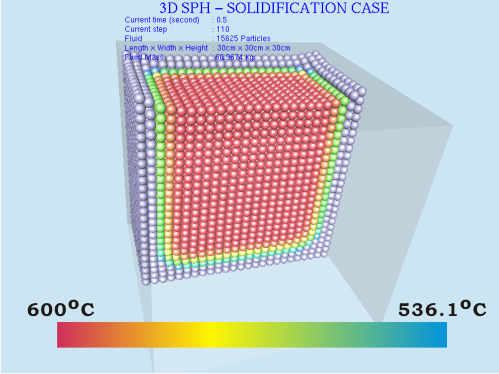}
               \label{fig:Result2}
       }
       \; 
       \subfigure[]{
               \centering
               \includegraphics[width=1.78in]{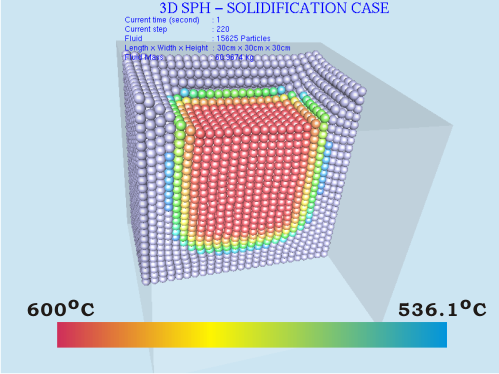}
               \label{fig:Result3}
       }
       \;      
	   \subfigure[]{
               \centering
               \includegraphics[width=1.78in]{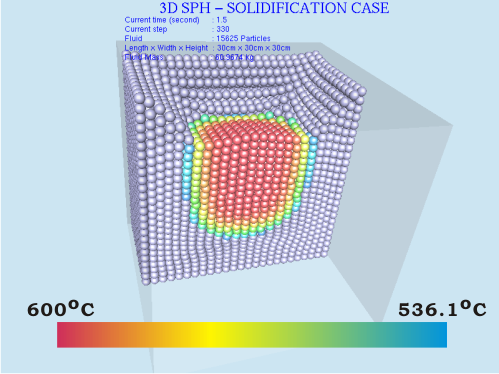}
               \label{fig:Result4}
       }
       \;      
	   \subfigure[]{
               \centering
               \includegraphics[width=1.78in]{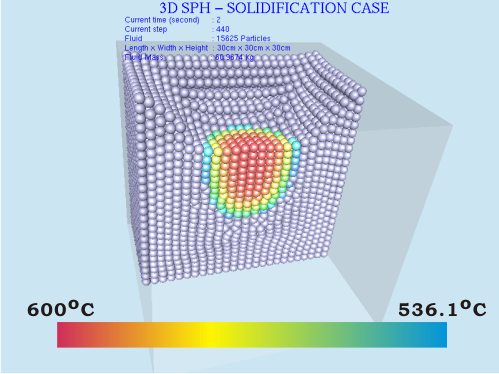}
               \label{fig:Result5}
       }
       \;      
	   \subfigure[]{
               \centering
               \includegraphics[width=1.78in]{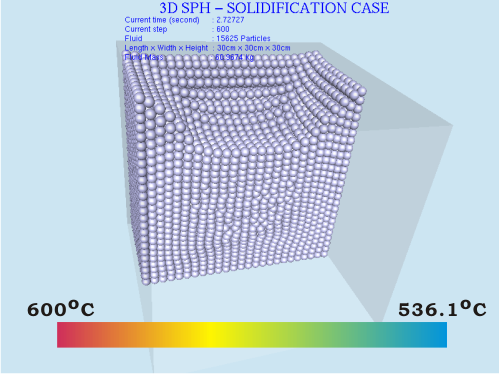}
               \label{fig:Result6}
       }
        
        \caption{Solidification process for several time step: (a) t = 0.0045 s, (b) t = 0.5 s, (c) t = 1 s, (d) t = 1.5 s, (e) t = 2 s, and  (f) t = 2.7272 s }\label{fig:simulationResult}
\end{figure}

In this simulation, we use 15625 liquid metal particles with 0.0106154 m of smoothing length. The simulation starts by arranging the liquid metal particles position in the well-ordered position. Then the particles are solidified rapidly through the side walls and the bottom wall. The results for several time steps are shown in the Figure \ref{fig:simulationResult}. Actually the domain is fully filled with the liquid metal particles. However, we display the results from the half part, so we can observe the inner part. The results show not only the temperature distribution but also the phase transition. The temperature distribution for the liquid phase and mushy phase are represented by the color bar. The red color represents the liquid phase with temperature above 589.7$^o$C and the blue color represents the lowest temperature for the mushy phase that is 536.1$^o$C. The solid phase is represented by silver color where the temperature is below 536.1$^o$C.

\begin{figure}[h]
\centering
\includegraphics[width=3in]
{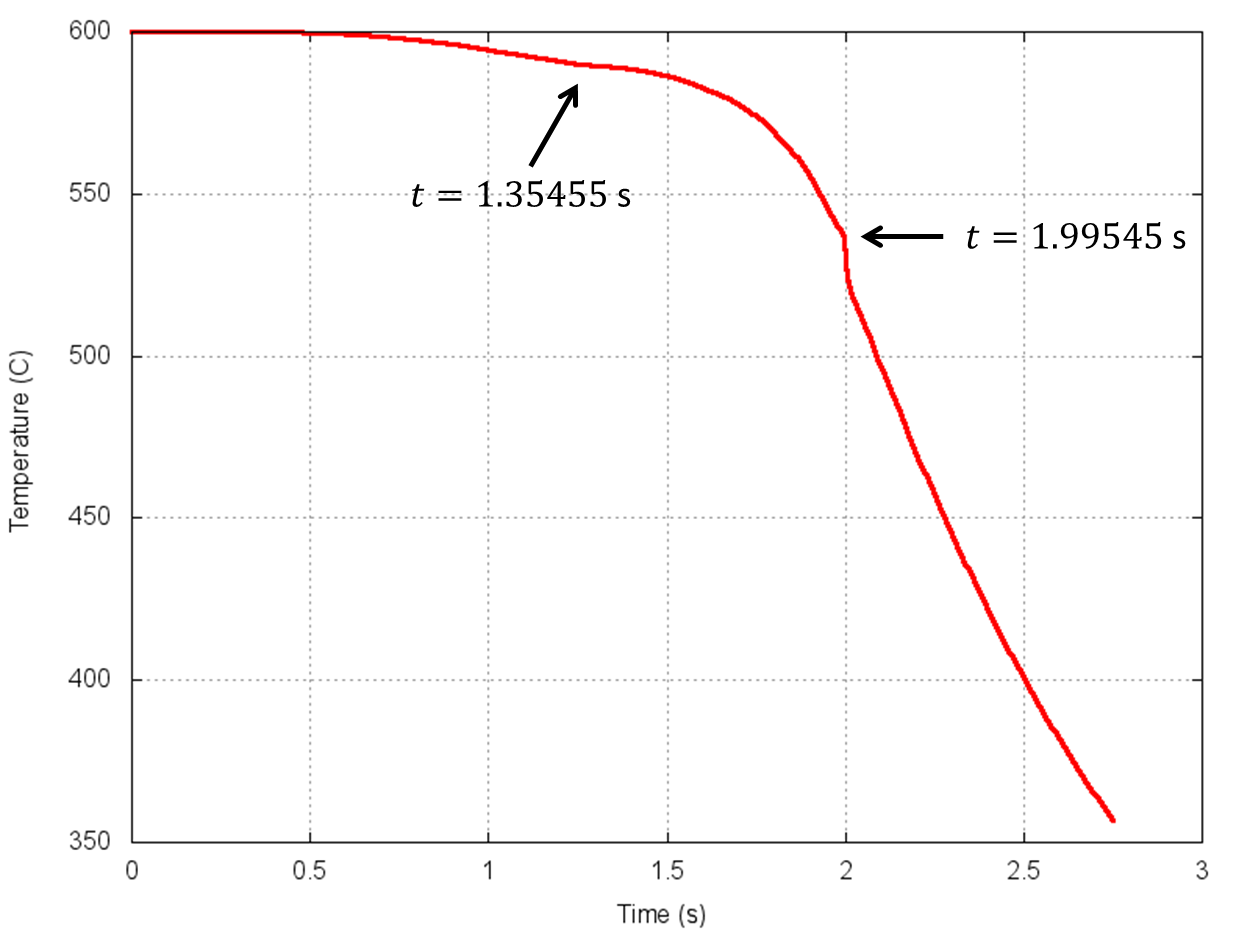}
\caption{Temperature history for a single particle. It reaches mushy phase at $t = 1.35455$ s and solid phase at $t = 1.99545$ s}
\label{fig:historyTemperature}
\end{figure}

The solidification process seems very fast. All of the liquid particles become solid after 2.7272 s. It is because we set the initial temperature of a single particle 600$^o$C while the initial temperature of the boundary is 200$^o$C. This value is very close to the liquidus temperature that is 589.1$^o$C. Another reason is that the thermal conductivities of mushy and solid phase are very high. So, if some particles are transformed to the solid phase, the heat transfer will be faster. As an example, the temperature histories of a single particle can be seen in Figure \ref{fig:historyTemperature}. It shows the different temperature drops at different phases. After the particle reaches the mushy phase or solid phase, the heat transfer works faster and leads the temperature to decrease rapidly. 

\begin{figure}[h]
       \centering
       \subfigure[]{
               \centering
               \includegraphics[width=1.78in]{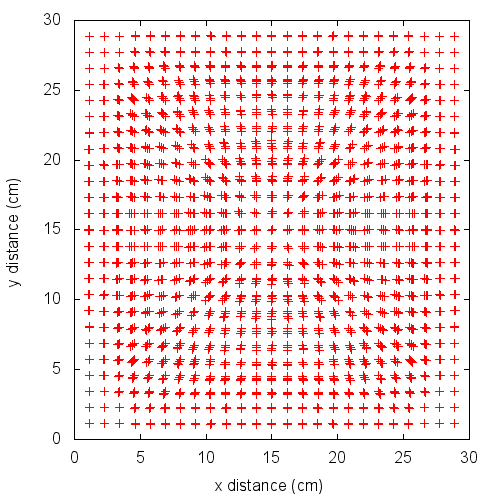}
               \label{fig:level1}
       }
       \;      
	   \subfigure[]{
               \centering
               \includegraphics[width=1.78in]{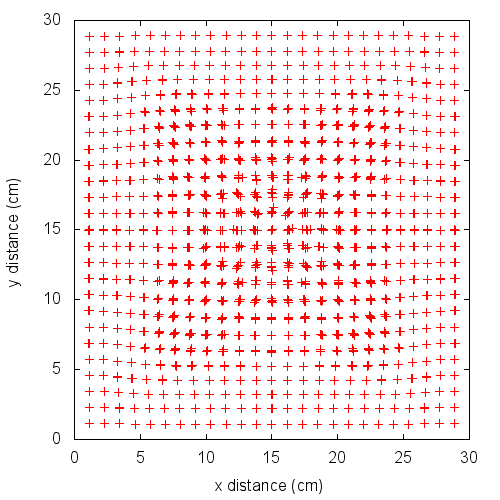}
               \label{fig:level2}
       }
       \;    
	   \subfigure[]{
               \centering
               \includegraphics[width=1.78in]{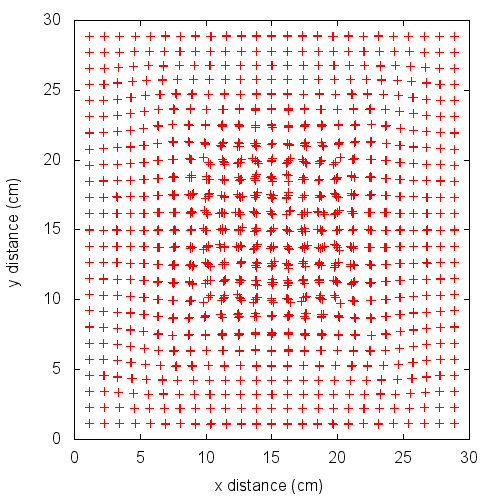}
               \label{fig:level3}
       }
        \caption{Density difference according to the number of particles at different layers: (a) 3 $\leq$ z $\leq$ 6 cm, (b) 6 $\leq$ z $\leq$ 9 cm, and (c) 9 $\leq$ z $\leq$ 12 cm }\label{fig:differentDensity} 
\end{figure}

From the results in Figure \ref{fig:simulationResult}, it can be seen that the volume of the liquid metal is decreasing every time step and it leads to the shrinkage cavity formation at the final solidification stage. It happened due to the difference in density where the solid phase has a higher density than the liquid phase. This affects the density of the mushy particles to continuously grow during the cooling down process. Hence the equation of states in Eq.(\ref{eq:EOS}) gives positive pressure; moreover the gravitational force also contributes to push the particles to the bottom. This simulation seems to approximate nicely the realistic phenomena. However, in the rapid cooling case, other defects such as cracks or voids often appear. Presence of cracks means there are some empty parts inside if we check the inner part of the particles distribution. The particles distribution for different layers can be seen in Figure \ref{fig:differentDensity}. We got different patterns on different layers. There are some lower densities and higher densities according to the number of particles. In the lower density part some cracks should appear. However, in our cases the cracks do not appear. It means that the model for liquid-solid interaction in 3D simulation should be improved. Our model which assumes the solid group as fluid with high viscosity and acts as undeformable object seems correct. It is because we can produce the shrinkage cavity formation and the simulation is stable until all the particle become solid. However, it is still not enough and needs to be improved such as by applying some particular forces between mushy and solid phases to produce other defects.
 
\section{Summary}
In this paper, the behavior of liquid metal solidification has been studied and simulated using smoothed particle hydrodynamics. The solidification process was done by installing the enthalpy method in SPH formulation to solve the heat transfer problem. The model of heat transfer process has been tested to static particle grid in one dimension and it gives a relatively good accuracy. The phase transition from liquid to solid was treated by considering the effect of latent heat and nonisothermal phase change to the model of heat transfer. The liquid-solid interaction model for 3D simulation also has been performed. The results showed the appearance of a defect when liquid metal was solidified. We got shrinkage cavity formation in the final solidification stage. However, in the rapid cooling process, other defects often appear, to obtain these defects, the model of liquid-solid interaction should be improved such as by adding a particular force between mushy and solid phases. The improvement of liquid-solid interaction model is left for the further study.

\end{document}